\documentclass[twocolumn, trackchanges]{aastex62}
\usepackage{amsmath}
\hypersetup{
	colorlinks	= true,
	linkcolor	= red,
	urlcolor	= cyan,
	citecolor	= blue
}

\turnoffedit

\graphicspath{{./images/}}

\newcommand{\taurex}{\mbox{$\mathcal{T}$-REx}}

\turnoffediting

\received{June 19, 2018}
\revised{May 16, 2019}
\accepted{May 18, 2019}
\published{June 25, 2019}

\submitjournal{ApJ}

\shortauthors{Damiano et al.}

\begin{document}
	
	\title{A Principal Component Analysis-based method to analyse high-resolution spectroscopic data}
	
	\correspondingauthor{Mario Damiano}
	\email{mario.damiano.15@ucl.ac.uk}
	
	\author[0000-0002-1830-8260]{M. Damiano}
	\affiliation{Department of Physics \& Astronomy, University College London, Gower Street, WC1E6BT London, United Kingdom}
	\affiliation{INAF - Osservatorio Astronomico di Palermo, Piazza del Parlamento 1, I-90134 Palermo, Italy}
	
	\author[0000-0002-9900-4751]{G. Micela}
	\affiliation{INAF - Osservatorio Astronomico di Palermo, Piazza del Parlamento 1, I-90134 Palermo, Italy}
	
	\author[0000-0001-6058-6654]{G. Tinetti}
	\affiliation{Department of Physics \& Astronomy, University College London, Gower Street, WC1E6BT London, United Kingdom}
	
	\begin{abstract}
		
		High-Resolution Spectroscopy (HRS) has been used to study the composition and dynamics of exoplanetary atmospheres.
		In particular, the spectrometer CRIRES installed on the ESO-VLT has been used to record high-resolution spectra in the Near-IR of gaseous exoplanets.
		
		Here we present a new automatic pipeline to analyse CRIRES data-sets. Said pipeline is  based on a novel use of Principal Component Analysis (PCA) and Cross-Correlation Function (CCF). The exoplanetary atmosphere is modelled with the \taurex\ code using opacities at high temperature from the ExoMol project. In this work we tested our analysis tools on the detection of CO and H$_{2}$O in the atmospheres of the hot-Jupiters HD209458b and HD189733b. The results of our pipeline are in agreement with previous results in the literature and other techniques.
		
	\end{abstract}
	
	
	\section{Introduction} \label{sec:intro}
	
	More than 4000 confirmed exoplanets are currently listed in the catalogues, together with basic planetary, stellar and orbital parameters as they become known. Transit and direct imaging spectroscopy from space and ground facilities are enabling the study of the physical and chemical properties of some of these exoplanets. 
	From space, one can observe exoplanet spectra in the UV, VIS and IR at low spectral resolution, without the hurdle of telluric contamination. Molecules, ions, atoms or absorbers able to imprint strong modulations in the recorded spectra, can be detected by using space-borne facilities, (e.g. \cite{Charbonneau2002, Linsky2010, Tinetti2007, Grillmair2008, Sing2016, Fraine2014, Damiano2017, Tsiaras2016B2016ApJ...832..202T, Tsiaras2016B2016ApJ...820...99T, Tsiaras2018}). 
	By contrast, observations from the ground at high-resolution (R$>$25,000) have enabled the detection of molecules or atoms whose weak absorptions are hard to detect at low spectral resolution. This is particularly true for alkali metals and CO which have been found  in the atmospheres of most hot-Jupiters analysed \citep{Redfield2008, Birkby2013, Birkby2017, Birkby2018, Brogi2014, Brogi2016, deKok2013, Snellen2010}. 
	
	High-resolution spectroscopy (HRS) allows to resolve molecular bands into individual lines. Using radial velocity measurements and techniques such as Cross-Correlation Function (CCF) we may separate three physically different sources: telluric absorption, stellar absorption and the planetary spectrum, which are normally entangled. 
	The aim -- but also the biggest challenge -- is to recognise the planetary signal among the telluric and the stellar signals, which can be orders of magnitude stronger. The  standard method used in the literature to analyse HRS data is to apply a number of corrections which involve the correction of the airmass, the subtraction of a modelled stellar spectrum from the data and the use of ad-hoc masks to eliminate residual strong features \citep{Birkby2013, Birkby2017, Birkby2018, Brogi2014, Brogi2016, Snellen2010}. 
	
	In this paper we present and assess an alternative automatic procedure to analyse HRS data \edit1{from the raw images to the final result}, which requires no manual intervention that could interfere with the objectivity and repeatability of the analysis. Our analysis method is based on a novel use of Principal Component Analysis (PCA) and Cross-Correlation Function (CCF). The exoplanetary atmosphere has been simulated using \taurex\ \citep{Waldmann2015B2015ApJ...802..107W, Waldmann2015B2015ApJ...813...13W} and line lists have been adopted from the ExoMol project \citep{Tennyson2016}.
	
	We applied our analysis method to two datasets recorded with VLT/CRIRES freely available on the ESO archive. The exoplanets observed are HD209458b and HD189733b (see Tab. \ref{tab:param}), the most studied planets up to date, and therefore good examples for testing new and/or different data analysis techniques. 
	HD209458b \citep{Mazeh2000} was the first planet analysed with high-resolution spectroscopy: \cite{Snellen2010} reported a detection of CO in its atmosphere. CO is absent in the Earth's atmosphere but also in the stellar spectrum due to the relatively hot temperature of HD209458. The  CO signal in the exoplanetary atmosphere should not be contaminated by the star and Earth's atmosphere.	
	By contrast the star hosting HD189733b is a K-type \citep{Bouchy2005}  showing CO absorption features in its spectrum: additional caution is therefore needed to remove the potential stellar contamination. \citet{Brogi2016} have reported the detection of H$_2$O and CO in the atmosphere of HD189733b. 
	
	In Section \ref{sec:method} we  describe our analysis method, in Section \ref{sec:results} we  show the results and in Section \ref{sec:discussion} discussion and conclusions are presented. 
	
	\section{Data Analysis} \label{sec:method}
	
\begin{figure*}[]
	\plotone{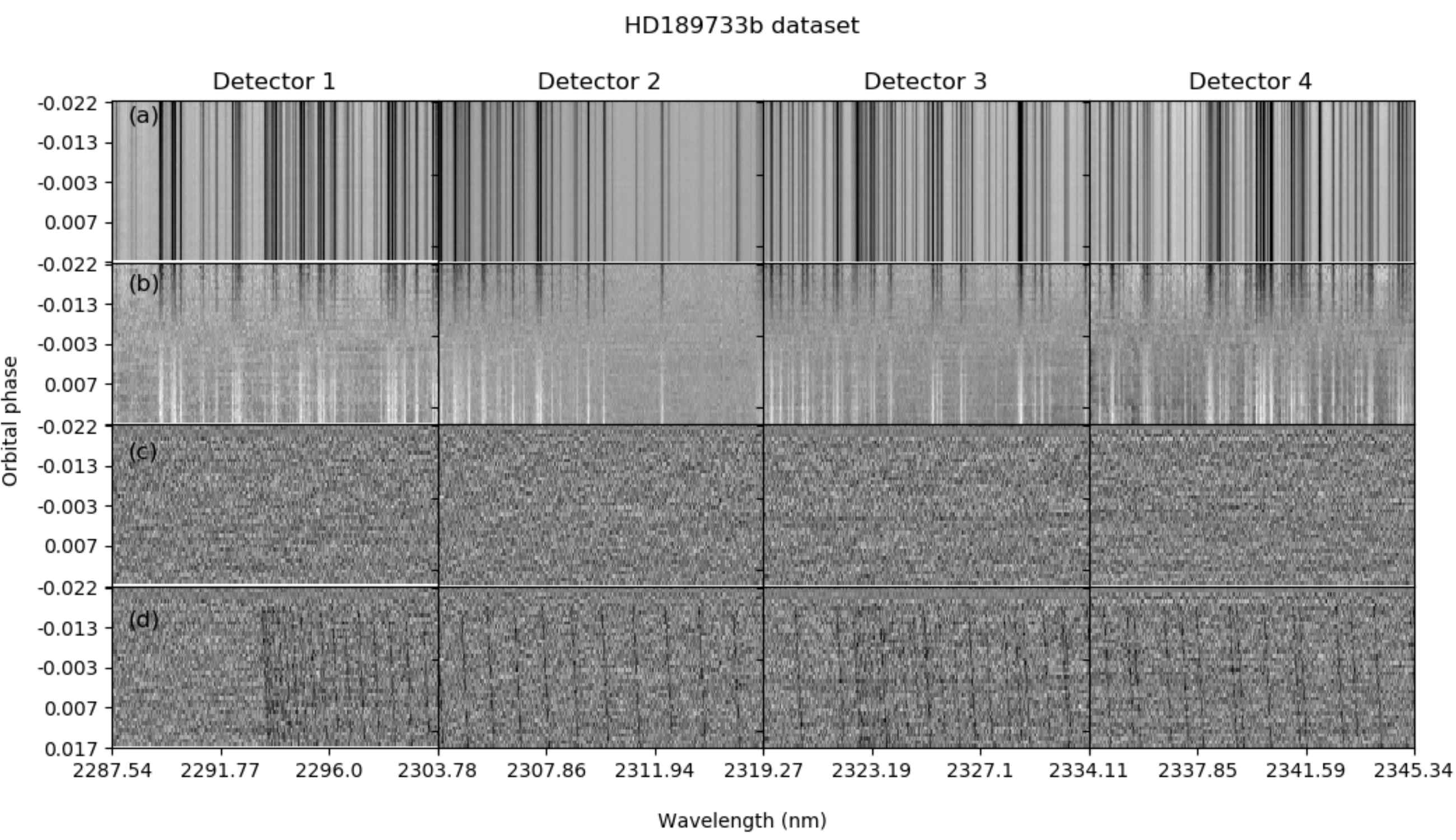}
	\caption{HD189733b dataset. In \textbf{(a)},  the data are shown after calibration, normalisation and spikes correction. In \textbf{(b)}, the data are shown after the median has been subtracted from each column. In \textbf{(c)}, the results of PCA are shown. In \textbf{(d)}, the data are shown after the application of PCA and the injection of the CO model. \label{fig:HD189}}
\end{figure*}

\begin{figure*}[]
	\plotone{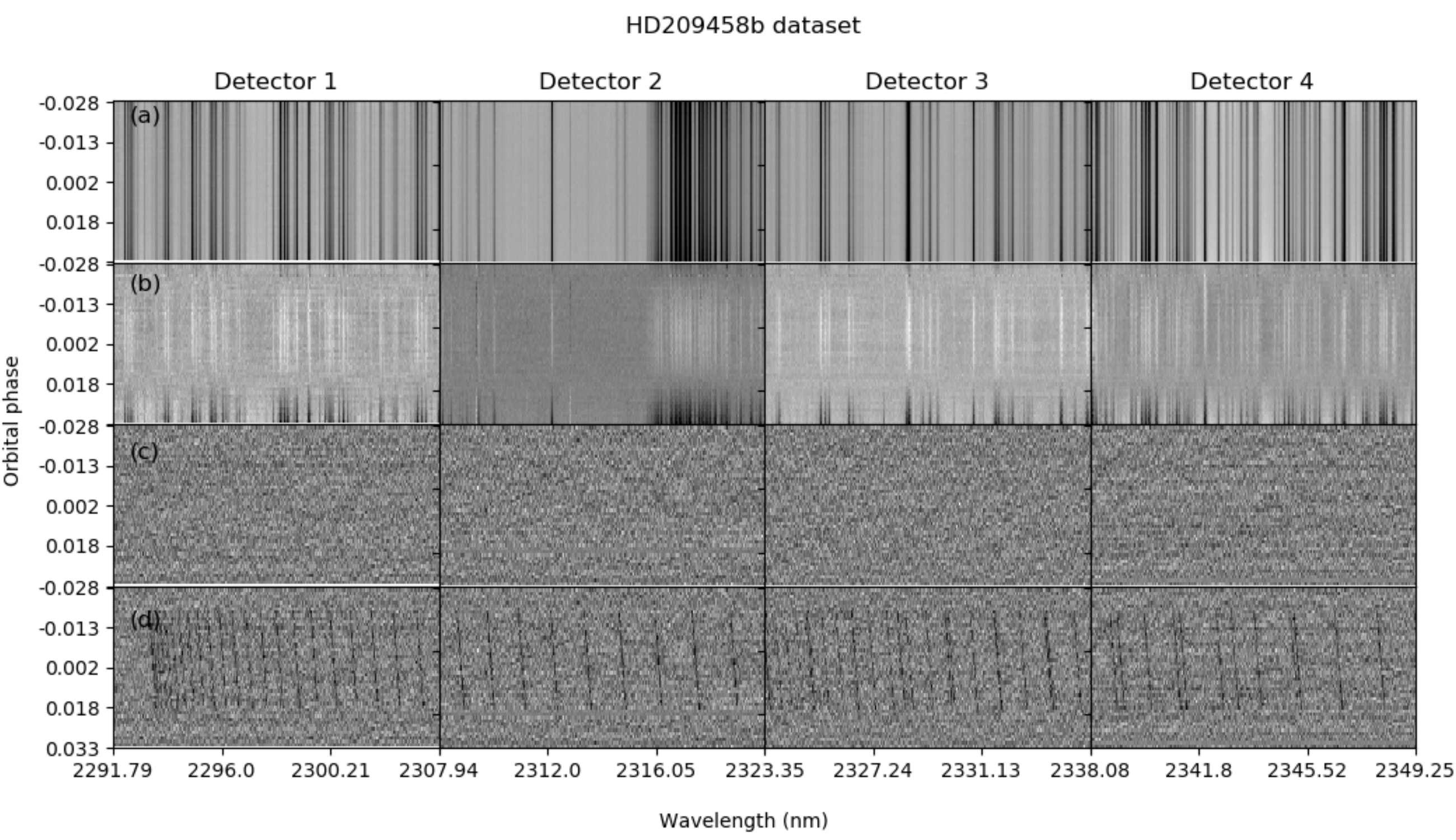}
	\caption{HD209458b dataset. In \textbf{(a)},  the data are shown after calibration, normalisation and spikes correction. In \textbf{(b)}, the data are shown after the median has been subtracted from each column. In \textbf{(c)}, the results of PCA are shown. In \textbf{(d)}, the data are shown after the application of PCA and the injection of the CO model. \label{fig:HD209}}
\end{figure*}

\begin{figure}[!h]
	\plotone{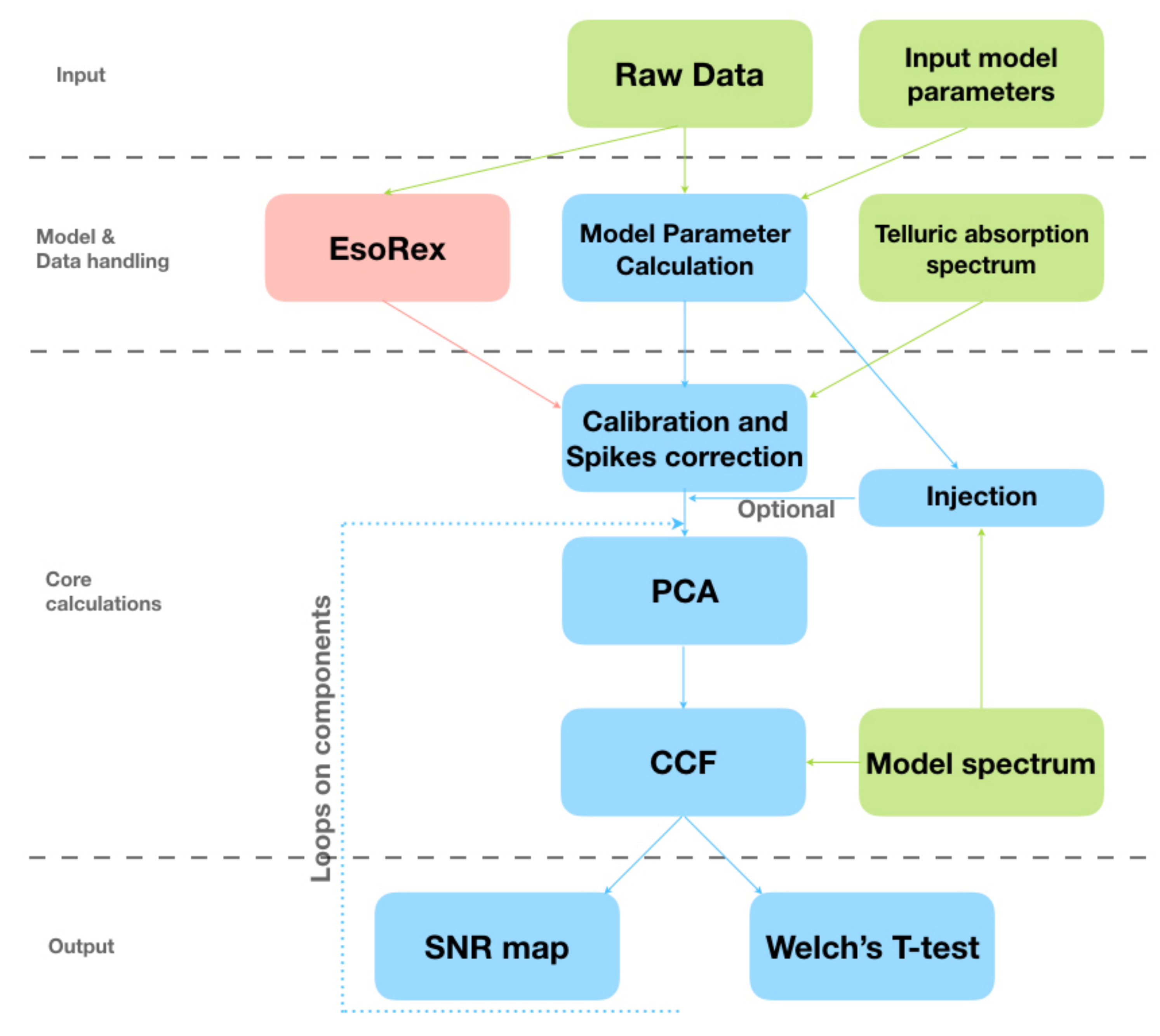}
	\caption{The box colours indicate different classes of action: green boxes represent an external input coming from other models or different sources (e.g user), the red box includes the external reduction algorithm. Finally, the blue boxes contain the calculations developed for the analysis of the data. \label{fig:workflow}}
\end{figure}

	We selected datasets relative to HD189733b and HD209458b which are publicly available on the ESO archive. These  are part of 289.C-5030(A) and 383.C-0045(A) programs (PI Snellen, I.) (Fig. \ref{fig:HD189} and \ref{fig:HD209} panels (a)). The observations have been recorded by using VLT/CRIRES at the highest resolution available (R$=100,000$) through the $0''.2$ slit. Both datasets cover a narrow wavelength range, i.e. $2287.54-2345.34\ $nm and $2291.79-2349.25\ $nm respectively, with three gaps ($\sim 200$ pixels per gap) due to the physical separation of CRIRES' detectors. Both datasets have been recorded with the nodding method ABBA for a better background subtraction \citep{Snellen2010, Brogi2016}. The steps of the analysis process are represented in Fig. \ref{fig:workflow} and they are described in following sections.
	
	\setcounter{table}{0}
	\begin{deluxetable}{c|cc}
		\tablecaption{Relevant parameters of the studied targets \label{tab:param}}
		\tablehead{
			\colhead{Parameter} 		& 			\colhead{HD189733} 								& 				\colhead{HD209458}												}
		\startdata
		Stellar Parameters 				&  															&  																			\\
		\hline
		$R_{\star}$ (R$_{\odot}$)			&			$0.756 \pm 0.018$			\tablenotemark{1}			&				($1.155_{-0.016}^{+0.014}$)			\tablenotemark{1}				\\
		$T_{eff}$ (K) 					&			$5040 \pm 50$				\tablenotemark{1}			&				$6065 \pm 50$						\tablenotemark{1} 				\\
		$M_{\star}$ (M$_{\odot}$)			&			$0.806 \pm 0.048$			\tablenotemark{1}			&				$1.119 \pm 0.033$					\tablenotemark{1}				\\
		$log(g_{\star})$ (csg)				&			$4.587 \pm 0.015$			\tablenotemark{1}			&				$4.361 \pm 0.008$					\tablenotemark{1}				\\
		$v_{sys}$ (kms$^{-1}$)			&			$-2.361 \pm 0.003$			\tablenotemark{2}			& 				$-14.7652 \pm 0.0016$				\tablenotemark{5} 				\\
		\hline
		Planet Parameters 				& 															&																			\\
		\hline
		$T_{eq}$ (K)					&			$(1201_{-12}^{+13})$ 		\tablenotemark{1}			&				$1449 \pm 12$						\tablenotemark{1}				\\
		$a$ (AU) 						&			$0.03120(27)$				\tablenotemark{3} 			& 				$(0.04707_{-0.00047}^{+0.00046})$		\tablenotemark{1} 				\\
		$R_p$ (R$_{Jup}$) 				&			$(1.178_{-0.023}^{+0.016})$	\tablenotemark{3}			&				$(1.359_{-0.019}^{+0.016})$			\tablenotemark{1}				\\
		$M_p$ (M$_{Jup}$)				&			$(1.144_{-0.056}^{+0.057})$	\tablenotemark{1}			&				$0.685 \pm 0.015$					\tablenotemark{1}				\\
		$P$ (days)					&			$2.21857567(15)$			\tablenotemark{4} 			& 				$3.52474859(38)$					\tablenotemark{6}				\\
		$T_0$ (BJD$_{UTC}$)			&			$2454279.436714(15)$		\tablenotemark{4}			&				$2452826.629283(87)$				\tablenotemark{6}				\\
		$I$ (deg)						&			$85.710 \pm 0.024$			\tablenotemark{4}			&				$86.71 \pm 0.05$					\tablenotemark{1}				\\ 
		\enddata
		\tablecomments{$^1$\cite{Torres2008}, $^2$\cite{Bouchy2005}, $^3$\cite{Triaud2009}, $^4$\cite{Agol2010}, $^5$\cite{Mazeh2000}, $^6$\cite{Knutson2007}}
	\end{deluxetable}

	\subsection{Data reduction and calibration}
	
	We adopted the pipeline provided by ESO (Crire kit Version-2.3.3) to process the raw data. The CRIRES' reduction pipeline has been embedded into our code thanks to ESO’s \textit{EsoRex} which is a command-line driven utility that can launch pipeline reduction routines (they are referred as recipes). These are individual scripts that perform specific actions to the input data. The reduction process performs the following steps:
	
	\begin{itemize}
		\item{dark subtraction;}
		\item{correction for detector non-linearity;}
		\item{flat-fielding;}
		\item{combination of nodding exposures;}
		\item{spectrum extraction;}
		\item{wavelength calibration.}
	\end{itemize}
	
	The master reduction files (e.g. dark and flat) are provided with the raw data, while the specific non-linearity correction files need to be downloaded from the archive\footnote{\url{https://www.eso.org/sci/facilities/paranal/instruments/crires/doc/VLT-MAN-ESO-14200-4032_v91.pdf}}. The 1D spectrum is extracted from the reduced images via an optimal extraction \citep{Horne1986}. By using the ABBA nodding method, we obtained 45 spectra for HD189733b and 51 for HD209458b dataset.
	
	To subtract and correct the telluric absorbtion, the calibration from the ESO pipeline is not accurate enough, we followed instead the procedure described in the literature \citep{Birkby2013, Birkby2017, Brogi2013, Brogi2014, Brogi2016, deKok2013, Snellen2010}, which involves a further calibration using the SKYCALC tool \footnote{\url{https://www.eso.org/observing/etc/bin/gen/form?INS.MODE=swspectr+INS.NAME=SKYCALC}}. This simulates the telluric absorption spectrum for a specific night. 
	
	The first step is to normalise each spectrum of each detector by dividing it by its median. This step is necessary to avoid  differences of baseline across spectra.
	After the normalisation, working on one detector at a time, we consider the mean spectrum. Here, the strongest lines (all the lines reaching a minimum $<0.8$) have been identified as homogeneously distributed as possible to cover the whole x-axis range. These same lines are also been identified within the telluric template. A Gaussian fit is then performed for each of these lines and the centroid is taken. The extracted spectrum centroids indicate the pixel number position. In the telluric template, instead, they indicate wavelength positions of the lines. We performed a fourth order polynomial fit to establish the relationship between pixels and wavelengths \citep{Snellen2010}. All the single spectra are then interpolated via a third order spline to the derived wavelength grid to have the same grid for all the spectra.
	
	We analysed each detector separately as a two-dimensional matrix, where the x-axis contains wavelengths and y-axis time: every row of this matrix is a spectrum, every column is a temporal-series at a given wavelength (see Fig. \ref{fig:HD189} and \ref{fig:HD209} panels (a)). We have therefore four different matrices. 
	Finally, the pipeline removes all the cosmic rays or spikes that could occur at the edges of the spectra due to the spline interpolation to the wavelength grid. The pipeline takes one column at a time of each 2D matrix, it calculates the median of the column and all the values outside $3\sigma$ from the median are set to the median value.
	
	\subsection{Decomposition Analysis (PCA)}
	
	The next steps involve the correction for telluric absorption, the subtraction of stellar signal and subtraction of correlated noise. \edit1{The use of an ad-hoc   mask  to remove the strongest telluric features has been frequently adopted in the literature (e.g. \cite{Snellen2010, Brogi2016}).}
	\edit1{Other works have considered an unsupervised linear transformation technique  to identify patterns in data, i.e. Principal Component Analysis (PCA). In \cite{Artigau2014} PCA was used to correct high-resolution spectra and improve the radial velocity accuracy for low-mass planetary detection.}
	\edit1{Similarly, in \cite{deKok2013}, \cite{RiddenHarper2016} and \cite{Piskorz2016, Piskorz2017}, PCA has been used to identify and de-trend the telluric absorption.} 
	\edit1{In those works  PCA  was used to decompose the data in the wavelength domain. }
	\edit1{Here, we explore the use of PCA applied to both wavelength and time domains. Additionally, we propose an objective criterion to determine an optimal selection of the principal components to be considered and the exact number of components to be subtracted.}
	More recently, the algorithm SYSREM developed by \cite{Tamuz2005} has been adopted to perform a similar task \citep{Birkby2017, Nugroho2017}. SYSREM allows to extract components iteratively one by one, however, the orthogonality of the extracted components is not guaranteed \citep{Tamuz2005}.
	
	As PCA is highly sensitive to data scaling,  we subtracted each column of the data matrices by its mean (Fig. \ref{fig:HD189} and \ref{fig:HD209} panels (b)).
	On a typical  spectroscopic dataset, the number of spectra are less than the wavelength bins, resulting in matrices that have more columns than rows. Here we adopt  the eigenvalue decomposition (EVD) of the covariance matrix \citep{Jolliffe2002}. 
	The dimension of the covariance matrix and the number of principal components (eigenvectors) are equal to the number of rows of the input matrix.  Two  cases are then considered:
	
		\begin{figure*}[]
		\plotone{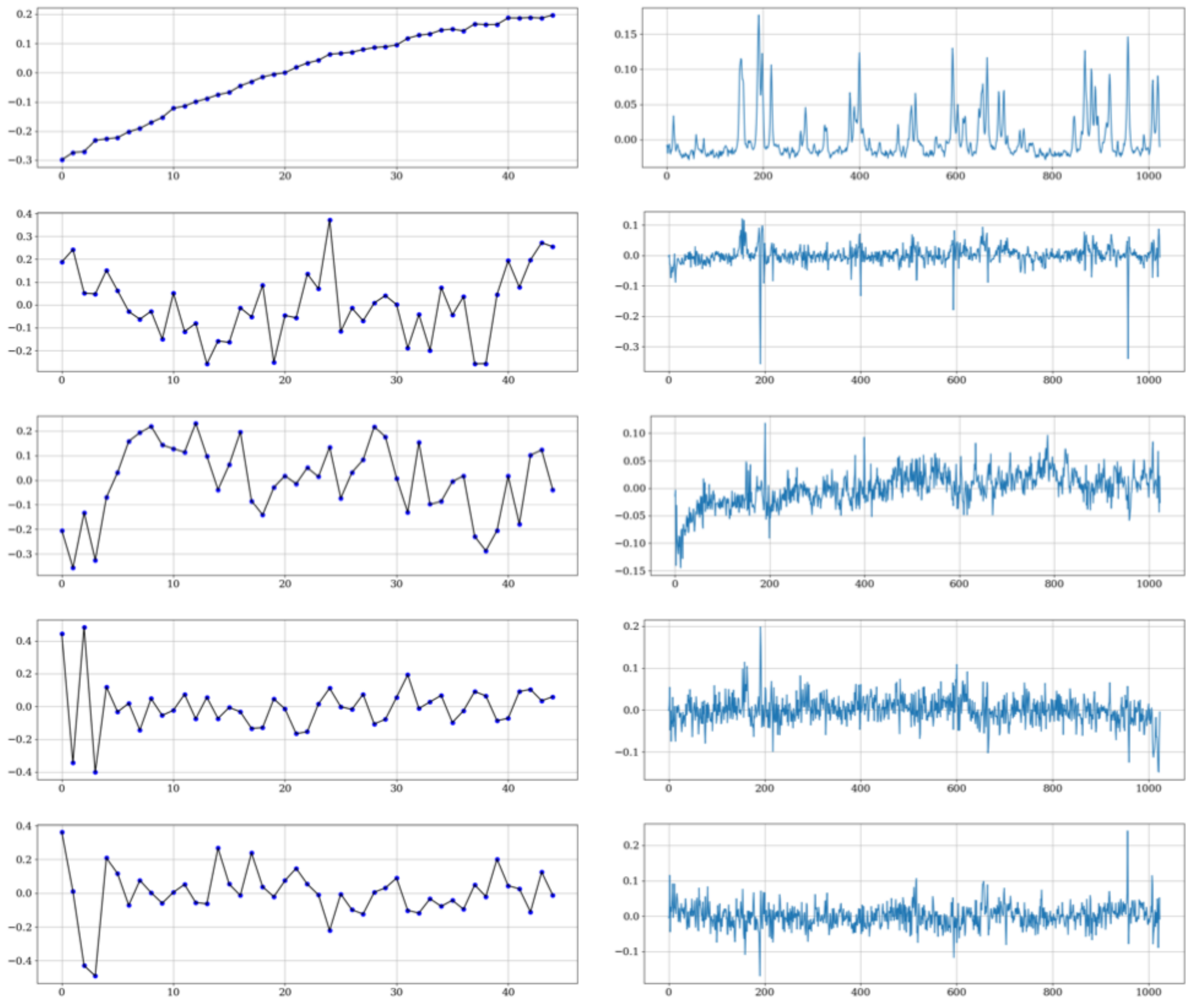}
		\caption{\textbf{Left panels}: first five eigenvectors of the TDM case.  \textbf{Right panels}:  first five eigenvectors of the WDM covariance matrix.  \label{fig:compo}}
	\end{figure*}
	
	\begin{itemize}
		\item{\textbf{time domain matrix (TDM)}; we use the individual spectra as rows  and the wavelength bins as columns.}
		\item{\textbf{wavelength domain matrix (WDM)}; we transpose the  matrix to have the spectra as columns and wavelength bins as rows;}
	\end{itemize}
	
\begin{figure*}[]
		\plotone{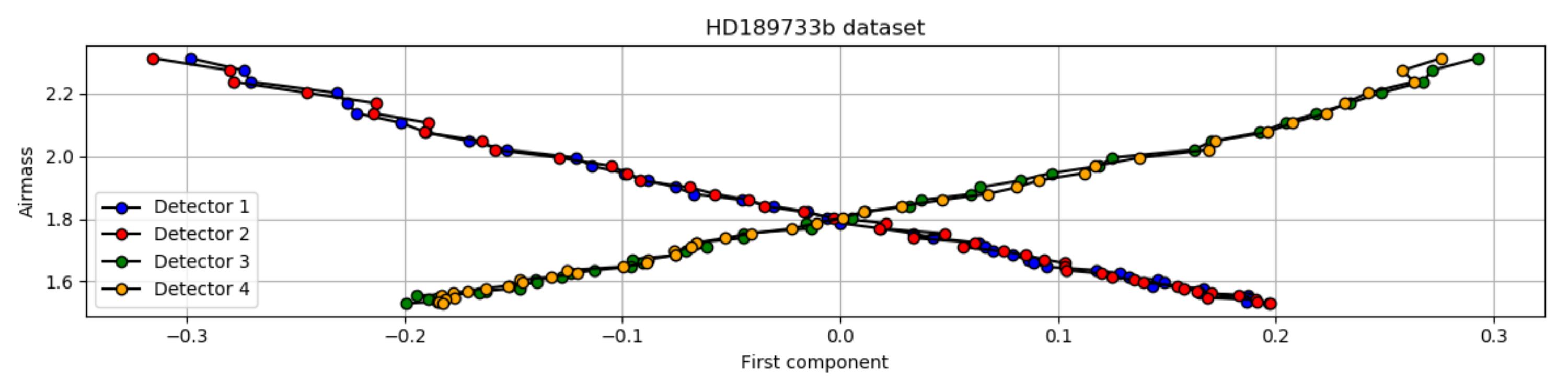}
		\caption{\edit1{Linear relation between the first component of each detector in the time domain and the recorded airmass for the HD189733b dataset.} \label{fig:comp-fis}}
	\end{figure*}

	In the WDM/TDM case the principal components (eigenvectors) contain the  information of the  correlations in the wavelength/time domain. 
	We consider, for example, the first detector of the HD189733b dataset:  Fig. \ref{fig:compo}  shows the first five components  of the TDM case (left) and the first five of the WDM (right).  The TDM components  contain  the time-domain information and the first one, in particular, is linked to the variations of the airmass: these are linearly correlated as we can appreciate from Fig. \ref{fig:comp-fis}. The WDM components  show the correlation in the wavelength domain and they appear to be  correlated with the telluric transmission spectrum. A good example is the strong feature around $200$ (Fig. \ref{fig:compo} x-axis unit, $\sim$2290 nm) that persists in all the components. 
	
	The TDM case has been chosen as best method for the following reasons:
	
	\begin{itemize}
		\item{the WDM component space cannot be fully described since there are more variables (1024 spectral bins) than observations (45 spectra for HD189733b and 51 for HD209458b dataset). The eigenvalues are  null after the 44$^{th}$ or 50$^{th}$ component, depending on the dataset;}
		\item{the application of a telluric mask is required if the WDM case is chosen to remove  most prominent telluric features that persist after PCA has been applied.}
	\end{itemize}
	
\begin{figure*}[]
		\centering
		\includegraphics[scale=0.5]{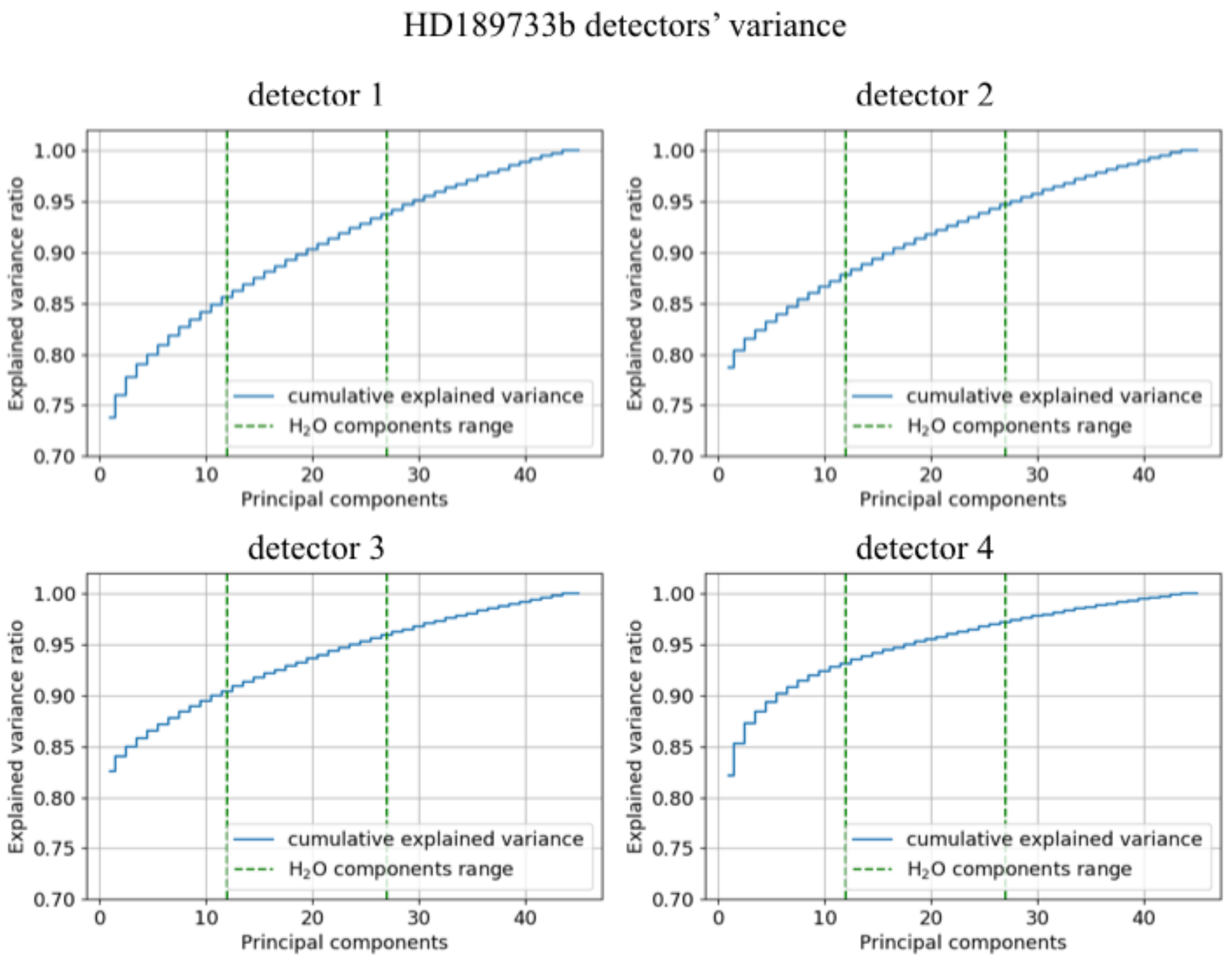}
		\caption{Detectors' variances of the PCA decomposition relative to the HD189733bb dataset. The first component always carries more than 75\% of the information. However, the variance is different for each of the detectors. \edit1{The green dashed lines indicate the calculated components range relative to the water vapour.} \label{fig:hd189_var}}
	\end{figure*}

\begin{figure*}[]
		\centering
		\includegraphics[scale=0.5]{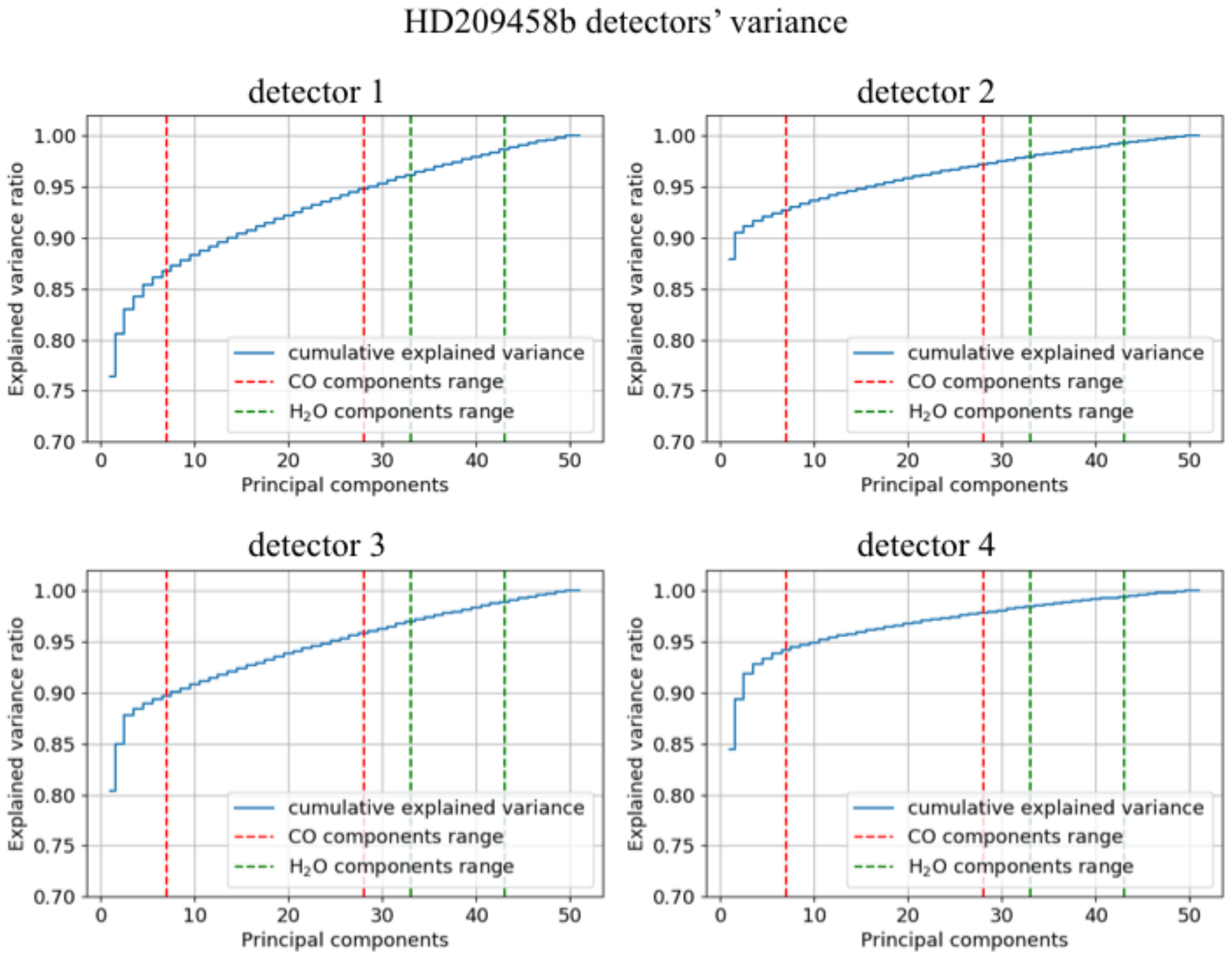}
		\caption{Detectors' variances of the PCA decomposition relative to the HD209458b dataset. The first component always carries more than 75\% of the information. However, the variance is different for each of the detectors. \edit1{The red dashed lines highlight the determined components range relative to the CO, while the green dashed lines are relative to the H$_2$O.} \label{fig:hd209_var}}
	\end{figure*}

	Following this choice, we calculated 50 TDM components for the HD209458b dataset and 44 components for the HD189733b dataset. They are equal to the number of recorded spectra minus one, due to the normalisation performed before the PCA decomposition.
	From the eigenvalues we calculated the \textit{explained variance ratio} (EVR) as follows: EVR$_{j}=\lambda_{j}/\sum\lambda_{i}$, where $\lambda_{i}$ are the eigenvalues. The EVR estimates the information carried by each principal component in  percentage.  The EVRs of each principal component for every detector of both datasets are shown in Fig. \ref{fig:hd189_var} and \ref{fig:hd209_var}. The first component has always the largest  variance ($\sim$ 80\%) as the telluric signal is the most significant.
	
	\edit1{The components are chosen  to maximise the S/N of the CCF peak expected at the theoretical $K_p$ and $v_{rest}$ of the planet (see. Eq. \ref{eq:kp}, Sec. \ref{SNR_matr}, Fig. \ref{fig:hd189_var} and \ref{fig:hd209_var}). This task is accomplished by removeing iteratively the low-order components, which supposedly are telluric or stellar in origin, and the high-order components, which account for non-correlated signal, presumably noise.
		The remaining components (time domain eigenvectors) are then projected back onto the original space.}
	
	After the application of PCA, each column of the output matrix was divided by its standard deviation to restore the S/N of the processed data \citep{deKok2013, Birkby2013, RiddenHarper2016, Nugroho2017} (Fig. \ref{fig:HD189} and \ref{fig:HD209} panels (c)).
	
	\subsection{Cross-Correlation Function (CCF)} 
	
	The cross-correlation function measures the similarity of two signals. It is also often called \textit{sliding dot product} since it returns a single value from the product of two signals when one slides over the other. Considering two series $\textit{\textbf{x}}$ and $\textit{\textbf{y}}$, the normalised cross-correlation $CCF$ at the delay $d$, for discrete series, is defined as follows \citep{Bracewell1965}
	
	\begin{equation}
	CCF(d) = \frac{\sum_{i} \left( \left( x(i) - \overline{x} \right) \cdot \left( y(i-d) - \overline{y} \right) \right)}{\sqrt{\sum_{i} \left( x(i) - \overline{x} \right)^2} \cdot \sqrt{\sum_{i} \left( y(i-d) - \overline{y} \right) ^2}}
	\end{equation}
	
	\noindent where $\overline{x}$ is the mean of the array $\textit{\textbf{x}}$, $\overline{y}$ is the mean of the array $\textit{\textbf{y}}$ and $i=0,1,2...N-1$.
	The idea of using such function is to find possible correlations between the data and an atmospheric model. The cross-correlation aims at matching similarities between the two signals. 
	
	The exoplanet atmospheric models have been simulated using \taurex\  \citep{Waldmann2015B2015ApJ...802..107W, Waldmann2015B2015ApJ...813...13W}. The CO and H$_2$O line lists at the planetary temperature were provided by ExoMol \citep{Tennyson2012, Tennyson2016}.
	
	Every row of the data matrix (every single spectrum), after the application of PCA, is cross-correlated with the simulated exoplanet atmospheric spectrum. This spectrum is interpolated to the same wavelength grid of the data, and it is then shifted from $-100$ to $100$ km/s with $1.0$ km/s as  step. The step is chosen based on the precision obtained during the calibration step ($\sim$ 1.0 km/s) and on the velocity resolution of the instrument (1.5 km/s). 
	
	The CCF transforms the matrices (one for each of the four CRIRES' detectors) from the wavelength domain to the velocity domain. The CCF matrices are then added together to obtain one single matrix (we will refer to it as CCF matrix).
	
\begin{figure*}[]
		\plotone{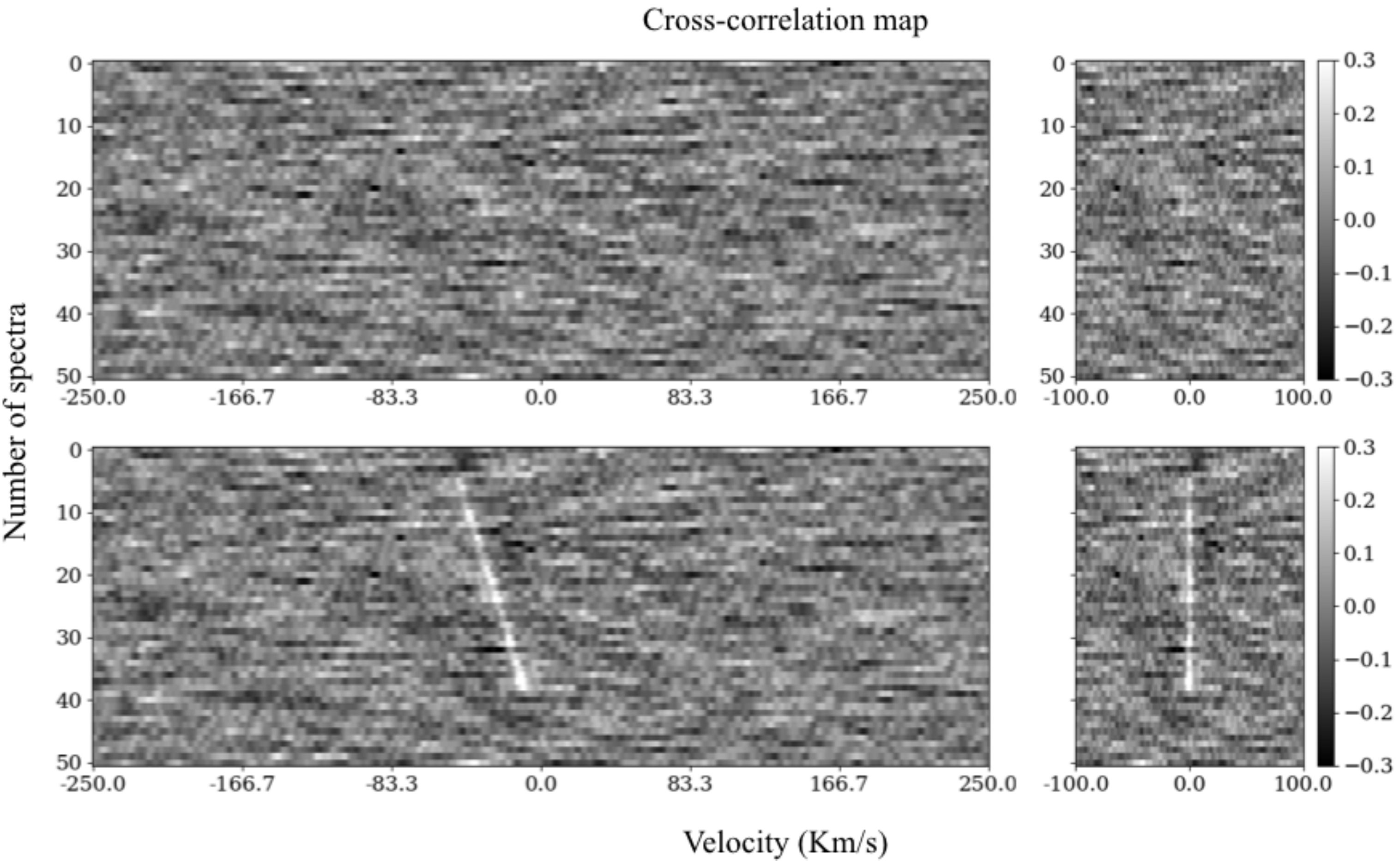}
		\caption{\textbf{Top left panel:} the four CCF of the four CRIRES' detectors summed together of the HD209458b dataset. \textbf{Bottom left panel:} same as top but with the model injected. The injection is $1\times$ the synthesised model ($R_p/R_{\star}$ $\sim10^{-3}$). \textbf{Top right panel:} cross-correlation after changing the reference frame from the Earth to the rest frame of the exoplanet. In this frame the planetary cross-correlation signal is aligned to zero kms$^{-1}$. \textbf{Bottom right panel:} same as top right panel but with the injection. The injected signal is aligned to zero kms$^{-1}$ in the exoplanet's rest frame. \label{fig:ccf}}
		\end{figure*}

\begin{figure*}[]
		\plotone{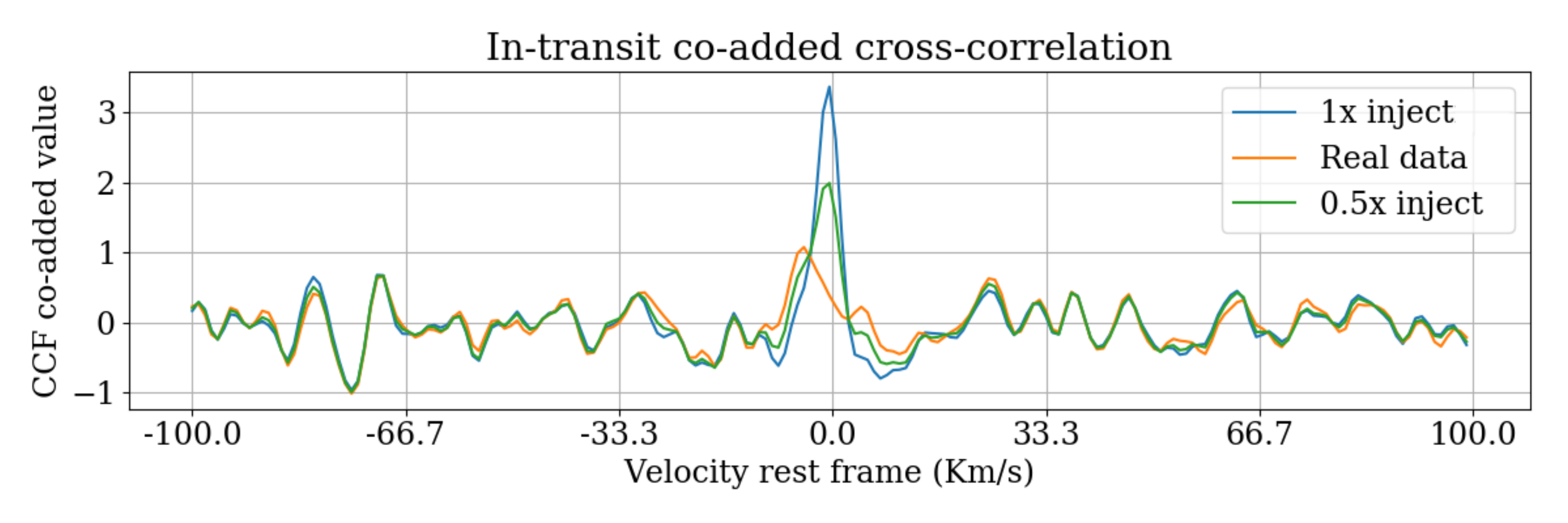}
		\caption{Cross-correlations of water vapour co-added in-transit for the HD209458b dataset. The injected signal and the planetary signal are still present after using PCA. The co-added CCFs are relative to HD209458b rest frame ($K_p = 145.041$ kms$^{-1}$). This graph has been generated considering PCA components from 33 to 43. \label{fig:H2O_hd209}}
	\end{figure*}

\begin{figure*}[]
	\plotone{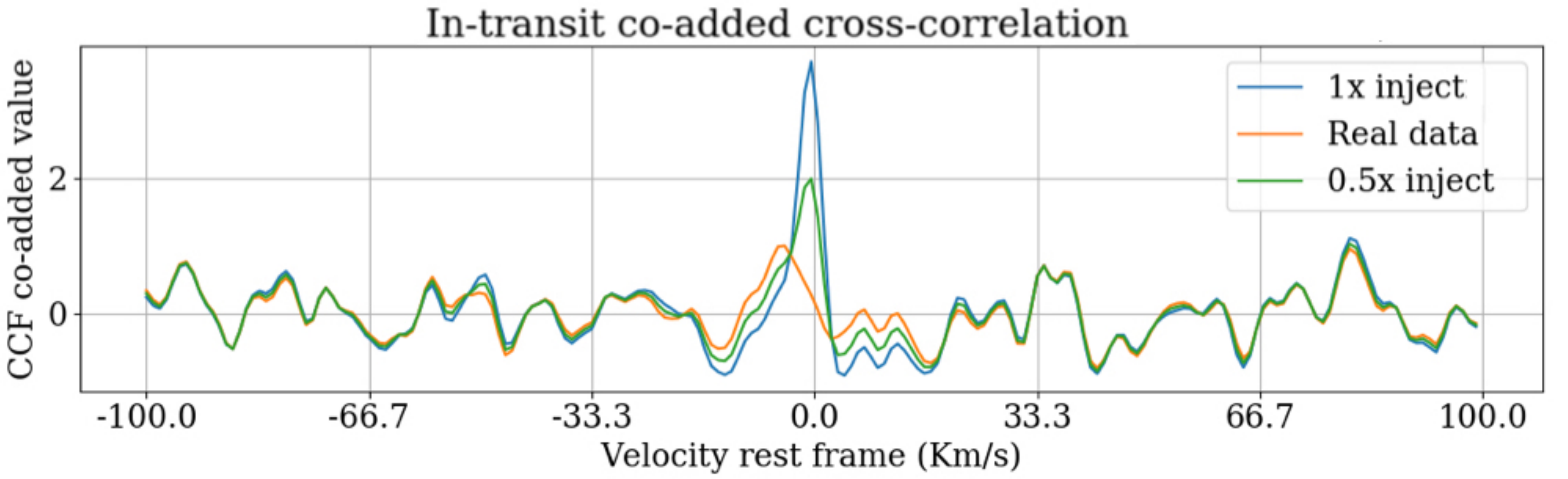}
	\caption{Cross-correlations of the planetary signal and of the injected water vapour co-added in-transit for the HD189733b dataset. The co-added CCFs are calculated at the theoretical orbital velocity of the planet HD189733b ($K_p = 152.564$ kms$^{-1}$). The CCFs are the result of the combination of the PCA components from 12$th$ to 27$th$. \label{fig:H2O_hd189}}
\end{figure*}

	At this stage the exoplanetary signal is not visible (see Fig. \ref{fig:ccf} top left panel).
	We then  injected  a synthetic signal with the orbital parameters of the planet (Fig. \ref{fig:ccf} bottom left panel) to predict the position of the signal (Fig. \ref{fig:ccf} bottom left panel) and to calculate the area of the S/N matrix interested by the planetary signal (see Sec. \ref{SNR_matr} and Fig. \ref{fig:H2O_hd209},  \ref{fig:H2O_hd189}). This step is performed after the calibration (before the PCA decomposition) and the effects of this process are not visible until the CCF is performed, because the signal intensity is at least three order of magnitude weaker than the telluric and stellar signals. The injected model cross-correlates with itself resulting, for example, in the signal shown in Fig. \ref{fig:ccf} bottom panels. 
	Finally, the injection process allows us to monitor the signal during the PCA decomposition. That helps to determine when the component subtraction  starts to erase part of the signal.
	
	\subsection{Signal extraction} \label{subsec:sig}v
	
	At this stage the planetary signal was barely visible or completely invisible, therefore we co-added the single CCFs (rows of the CCF matrix) in-transit to obtain the integrated signal from the planet. As the data were aligned to the telluric spectrum reference system, the planetary spectrum moved across time, we then re-aligned the single CCFs to the reference system of the  planet by computing the following correction:
	
	\begin{equation} \label{eq:vp}
	V_p\ =\ K_p\ sin\left[2\pi\phi(t)\right]\ + v_{sys}\ + v_{bary}(t).
	\end{equation}
	
	\noindent where $K_p$ is the radial velocity amplitude of the planet (Eq. \ref{eq:kp}) and $\phi(t)$ is the orbital phase (Eq. \ref{eq:phase})
	
	\begin{equation} \label{eq:kp}
	K_p\ =\ v_{orb}\ sin(i)
	\end{equation}
	
	\begin{equation} \label{eq:vorb}
	v_{orb}\ =\ \frac{2\pi a}{P_{orb}}
	\end{equation}
	
	\begin{equation} \label{eq:phase}
	\phi(t)\ =\ \frac{t-T_0}{P_{orb}}
	\end{equation}
	
	\noindent All parameters are listed in Table \ref{tab:param}.
	
	Once all CCFs were aligned to the planetary rest frame, we co-added in time only the in-transit CCFs. These were selected by computing the transit time \citep{Seager2003, Kipping2010}. When all the in-transit cross-correlations are summed together, the 2D cross-correlation matrix is reduced to a 1D signal, which is connected to the theoretical orbital velocity of the planet. To explore different orbital velocities we proceeded as follows:
	
	\begin{enumerate}
		\item{we let $K_p$ varying from $0$ to $250$ km/s with $1$ km/s step;}
		\item{for each $K_p$ we applied the correction in Eq. \ref{eq:vp} to every single CCF in the CCF matrix;}
		\item{we summed only the in-transit cross-correlations.}
	\end{enumerate}
	
	In this way we were able to explore all possible orbital velocities including those corresponding to the host star. Following the previous steps, we obtained a matrix with $K_p$ on y-axis and velocity rest frame along x-axis ($v_{rest}$). From this matrix two different outputs were extracted: the S/N map and the T-test statistic.
	
	\subsection{S/N matrix} \label{SNR_matr}
	
	We considered the last matrix obtained, i.e. $K_p$ on y-axis and $v_{rest}$ on x-axis. We calculated the standard deviation of this matrix excluding those points potentially correlated to the planetary signal ($|v_{rest}|\ <\ 15\ $km/s) and we divided the entire matrix by this value. We refer to the obtained matrix  as the S/N matrix (Fig. \ref{fig:hd209_result} and \ref{fig:hd189_result} left panels).
	
	To assign an uncertainty to the $K_p$ value we followed the same procedure as reported in \cite{Brogi2016}: i.e. we took the maximum value of the matrix and, fixing the relative $v_{rest}$, we calculated the $K_p $ interval where the S/N dropped by a unit around the $K_p $ peak. The same approach was used to determine the uncertainty for the $v_{rest}$.
	
	The S/N map is not only useful to represent visually the results but also to inspect whether spurious signals or telluric residuals are present. These signals may have high S/N value but are located at different $K_p$ and/or $v_{rest}$  from those expected for the planetary signal.
	
	We calculated the S/N matrix for each excluded principal component. Two loops need to be performed to explore the entire principal component space: the first loop subtracts higher variance components onwards and aims to remove the most correlated signal (e.g. telluric absorption and stellar signal). The second loop subtracts lower variance components backwards and aims to remove uncorrelated noise from the data. Finally, the principal components were selected to maximise the peak of the S/N matrix in correspondence of the expected planetary $K_p$ and $v_{rest}$.
	
	\subsection{Welch' T-test statistics} \label{sec:t-test}
	
	The Welch's T-test is used to test the hypothesis that two populations have equal means. This test compares the population of points on the CCF map connected to the planetary signal with those that are not.
	
	From the CCF matrix we defined, as done in the literature \citep{Brogi2016, Nugroho2017}:
	
	\begin{itemize}
		\item{\textit{in-trail}, those values inside a squared box centred on the CCF' peak with a radius of $\pm15$ km/s;}
		\item{\textit{out-trail}, those values outside the \textit{in-trail} box}
	\end{itemize}
	
	We extracted two families of values from the CCF matrix and these were compared through Welch's T-test (Fig. \ref{fig:hd209_result} and \ref{fig:hd189_result} right panels). The test, calculated using \textit{scipy.stats.ttest\textunderscore ind} in \textit{python}, provides a \textit{p-value} (two-tailed) which was converted into \textit{$\sigma$-value} (significance interval) through the inversion of the \textit{survival function} (SF)
	
	\begin{equation}
	\sigma_{value}\ =\ SF^{-1}(p\text{-}value\ /\ 2) 
	\end{equation}
	
	\noindent where the $SF^{-1}$ is the inverse of the survival function that is calculated from the \textit{cumulative density function} (CDF) as follows:
	
	\begin{equation}
	SF\ =\ 1\ -\ CDF.
	\end{equation}
	
	\section{Results} \label{sec:results}
	
	\begin{description}
		\item[HD209458b]{for the cross-correlation process, we assumed an isothermal $T-p$ profile at $T= 1400$K, with the pressure varying from $10^{-5}$ to $10^4$ Pa. We did not include clouds nor line-broadening due to the rotation of the planet. We used $10^{-3}$ as Volume Mixing Ratio (VMR) for both molecules, this value is compatible with chemical models' predictions for Hot-Jupiters atmospheres \citep{Venot2012}. The same value was also used by \cite{Snellen2010} for the CO.
			
			The signal obtained for CO peaks at S/N=5.7 (Fig. \ref{fig:hd209_result} top left panel and Tab \ref{tab:results}). The signal is compatible with the planetary orbital parameters ($K_p = 148_{-15} ^{+16}$ km/s, $v_{rest} = -3.0_{-1.1}^{+1.3}$ km/s). This result has been obtained by considering components from the 7$th$ to the 28$th$ \edit1{(Fig. \ref{fig:hd209_var} red lines)}. 
			The statistical significance of the result is also confirmed by the Welch's T-Test (Fig. \ref{fig:hd209_result} top right panel). Using a box of radius 15 km/s the null hypothesis is rejected with a confidence greater than 7$\sigma$, the shift of the \textit{in-trail} population is noticeable with respect to the \textit{out-trail} values that are, instead, distributed as a Gaussian centred to zero.
			
			The signal of the water vapour is more difficult to detect since the Earth's atmosphere also contains water. To extract the planetary signal a robust telluric correction is required, and therefore several components need to be subtracted using PCA. A signal at the compatible planetary parameters is observable in the S/N map in Fig. \ref{fig:hd209_result} bottom left panel. The maximum peaks at S/N=3.95, $K_p = 140 _{-16} ^{+25}$ km/s and $v_{rest} = -4.0_{-1.6}^{+1.4}$ km/s and it is obtained considering components from the 33$th$ to the 43$th$ \edit1{(Fig. \ref{fig:hd209_var} green lines)}. To demonstrate that the H$_2$O planetary signal survives after 33 components have been subtracted, Fig. \ref{fig:H2O_hd209} shows the in-transit co-added cross-correlation relative to the range of components aforementioned. Both the injected and non-injected signal survive to the PCA correction (note that the injected signal does not include any atmospheric dynamics, so it is not blue-shifted as the planetary signal). Moreover, the co-added cross-correlation value is lower with respect to the CO case meaning that the concentration of water is lower than CO or that PCA has erased part of the signal.
			Finally, the Welch's T-Test is performed on the \textit{in-trail} and \textit{out-trail} populations (Fig. \ref{fig:hd209_result} bottom right panel and Tab \ref{tab:results}). In this case the shift of the \textit{in-trail} population is not as strong as in the CO case but the null hypothesis is rejected with a confidence greater than 6$\sigma$.}
		
		\begin{deluxetable}{c|cc}
			\tablecaption{This work and previous results \label{tab:results}}
			\tablehead{
				\colhead{Parameter} 		& 			\colhead{HD189733} 								& 				\colhead{HD209458}												}
			\startdata
			\hline
			Previous results			 	& 			\cite{Brogi2016}									& 				\cite{Snellen2010}												\\
			\hline
			S/N$_{CO}$					&			-												&				-															\\
			$K_{p,\ CO}$ (kms$^{-1}$)		&			$205_{-51}^{+38}$									&				-															\\
			$v_{p,\ CO}$ (kms$^{-1}$)			&			-												&				$140 \pm 10$													\\
			$v_{rest,\ CO}\ $(kms$^{-1}$)		&			$-1.6_{-1.8}^{+2.0}$									&				$\sim$ 2														\\
			S/N$_{H{_2}O}$				&			5.5												&				-															\\
			$K_{p,\ H{_2}O}$ (kms$^{-1}$)		&			$183_{-59}^{+38}$									&				-															\\
			$v_{rest,\ H{_2}O}\ $(kms$^{-1}$)	&			$-1.58_{-1.50}^{+1.65}$								&				-															\\
			\hline
			Results						& 			This Work											&				This Work														\\
			\hline
			S/N$_{CO}$					&			5.24												&				5.7															\\
			$K_{p,\ CO}$ (kms$^{-1}$)		&			$190	 \pm 16$										&				$148_{-15}^{+16}$												\\
			$v_{rest,\ CO}\ $(kms$^{-1}$)		&			$-3.0_{-1.3}^{+1.0}$									&				$-3.0_{-1.1}^{+1.3}$												\\
			W T-Test$_{CO}$ ($\sigma$)		&			-												&				21.62														\\
			S/N$_{H{_2}O}$				&			3.69												&				3.95															\\
			$K_{p,\ H{_2}O}$ (kms$^{-1}$)		&			$167_{-21}^{+32}$									&				$140_{-16}^{+25}$												\\
			$v_{rest,\ H{_2}O}\ $(kms$^{-1}$)	&			$-4.0_{-1.8}^{+2.0}$									&				$-4.0_{-1.6}^{+1.4}$												\\
			W T-Test$_{H{_2}O}$ ($\sigma$)	&			5.21												&				6.56															\\
			\enddata
		\end{deluxetable}
	
			\begin{figure*}[]
		\plotone{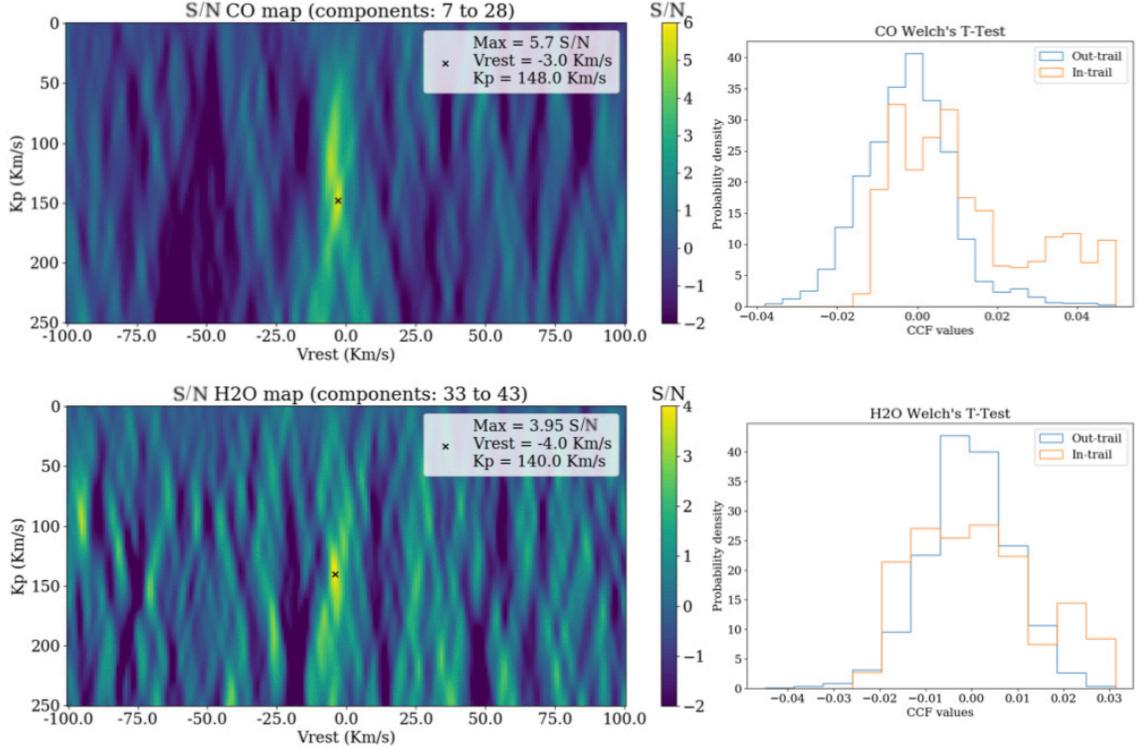}
		\caption{Results for the HD209458b dataset. \textbf{Top left panel:} S/N map for the carbon monoxide. The maximum point is compatible with the planetary orbital parameters. \textbf{Top right panel:} distributions (i.e. \textit{in-trail} and \textit{out-trail}) used to compute the Welch's T-Test. The null hypothesis is rejected with a confidence greater than 7$\sigma$. \textbf{Bottom left panel:} S/N map of the water vapour. The peak is compatible with the planetary parameters. \textbf{Bottom right panel:} distribution used to compute the Welch's T-Test. The null hypothesis is rejected with a confidence of 6.56$\sigma$. \label{fig:hd209_result}}
	\end{figure*}
		
		\item[HD189733b]{the planetary transmission spectrum was modelled with an isothermal $T-p$ profiles at $T= 1000$K. The pressure varies from $10^{-5}$ to $10^4$ Pa and we did not include clouds or any lines broadening due to the rotation of the planet. We used $10^{-3}$ as Volume Mixing Ratio (VMR), this value is compatible with chemical models' predictions for Hot-Jupiters \citep{Venot2012}.
			
			The CO detection is highly difficult since the star, being a K-type star (T$\sim4900$ K), contains CO in the outer regions. In \cite{Brogi2016} a master stellar spectrum has been simulated and subtracted to the data, but the stellar contamination continued to be persistent also in the result. In this work, PCA was not as effective as in the HD209458b case because the star spectrum moves 1-2 pixels on the detector preventing an optimal correction. The result (see Fig. \ref{fig:hd189_result} top left panel and Tab. \ref{tab:results}) is compatible with the one claimed by \cite{Brogi2016} (S/N=5.1, $K_p=194_{-41}^{+19}$ km/s, $v_{rest} = -1.7_{1.2}^{1.1}$ km/s), however the error on the $K_p$, being smaller than the one reported in literature, does not include the theoretical value of the orbital velocity of the planet ($K_p = 152.564$ kms$^{-1}$). The signal determined at lower $K_p$ ($\sim85$ km/s) is due to stellar contamination, that results from the Rossiter-McLaughlin effect combined with the change of reference frame from the Earth to the barycentric one \citep{Brogi2016}.
			
			Concerning water vapour, the same discussion done for HD209458b can be applied here. The planetary water signal needs to be disentangled from the telluric absorption.
			The result obtained \edit1{(Fig. \ref{fig:hd189_var} green lines and Fig. \ref{fig:hd189_result})} is compatible with both the literature and the theoretical parameters, e.g. see Fig. \ref{fig:H2O_hd189} where the planetary signal is compared with the injected one. The injected signals do not account for $v_{rest} \neq 0$ km/s, we can appreciate the data being blue-shifted. 
			Here the Welch's T-Test confirms that the null hypothesis can be rejected with a confidence greater than 5$\sigma$ (Fig. \ref{fig:hd189_result} bottom right panel and Tab \ref{tab:results}).}
	\end{description}
	
		\begin{figure*}[]
		\plotone{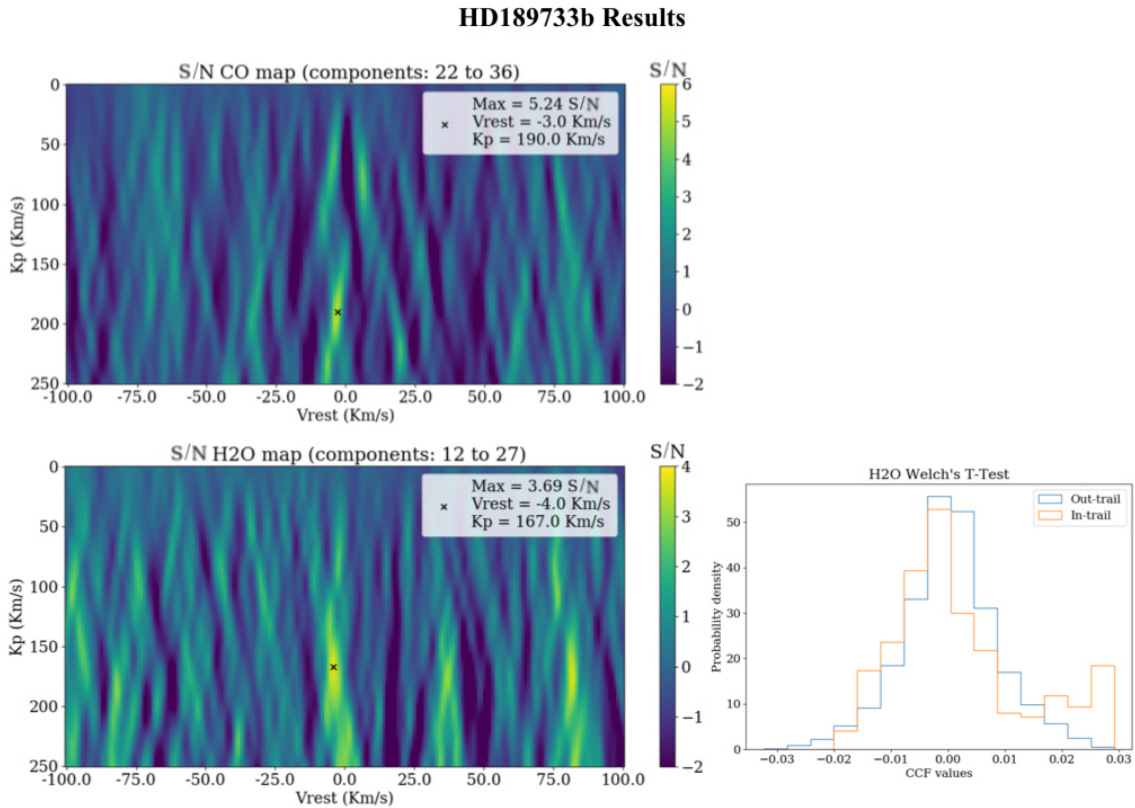}
		\caption{Results for the HD189733b dataset. \textbf{Top left panel:} S/N map for the carbon monoxide. The maximum point is compatible with the result reported in \cite{Brogi2016} but it is not compatible with the expected value. \textbf{Bottom left panel:} S/N map of the water vapour. The peak is compatible with the planetary parameters. \textbf{Bottom right panel:} distribution used to compute the Welch's T-Test. The null hypothesis is rejected with a confidence of 5.21$\sigma$. \label{fig:hd189_result}}
	\end{figure*}
	
	We performed an additional  test by cross-correlating the telluric model used in the calibration process with the data, to check if any telluric signal still persists. Using the components  reported in the results we did not notice any significant correlation with the telluric signal at the position of the planetary parameters. 
	\edit1{We have also tried to cross-correlate other molecules with the data (e.g. CH$_4$, NH$_3$ and CO$_2$) but no correlations have been found.}
	
	\section{Discussion and conclusions} \label{sec:discussion}
	We presented here and tested a new automatic method\edit1{, from the raw images to the final result,} based on the iterative use of PCA and CCF to re-analyse two CRIRES datasets observed with high resolution spectroscopy technique. \edit1{ Our pipeline does not assume prior knowledge, e.g. the variation of the airmass, nor does require ad-hoc corrections, e.g. masks to remove telluric lines. The PCA components are automatically selected by maximising the signal extracted. The algorithm is able to calculate the final result (S/N maps and W T-Test) without  manual intervention, allowing to analyse rapidly many data sets.}
	
	CO and H$_2$O have been detected in the HD209458b dataset, and H$_2$O in the HD189733b dataset. 
	The detection of CO in the HD209458b atmosphere is supported by an S/N peak of 5.7 at $K_p$ and $v_{rest}$ compatible with the planetary orbital parameters. Contrary to CO, H$_2$O is present in the Earth's atmosphere and therefore an accurate telluric correction is required. The lower S/N peak may be due to a lower concentration of H$_2$O with respect to CO in the atmosphere of HD209458b, or part of the signal might have been removed by PCA. In both detections a blueshift has been observed and this could be explained with high altitude winds. The results presented here are in agreement with the results published by \cite{Snellen2010}.
	
	Concerning HD189733b, using our method, we have been able to detect H$_2$O.  Even in this dataset a blueshift of the signal has been observed and also in this case it could be associated with high altitude winds. The detected CO signal, is compatible with the literature \citep{Brogi2016} but it is not in agreement with the theoretical radial velocity of the planet, and could be due to stellar contamination (K-type star \edit1{shows CO spectral features}).
	
	\edit1{We note that the requirement on maximisation of the S/N peak may lead to a biased $K_p$ and $v_{rest}$ values. In the work presented here this effect, if present, does not exceed the reported errorbars:   changes of less than one pixel are found between one component and the others (one pixel corresponds to the CCF step).}
	
	We note that the EVR is different for each detector (Fig. \ref{fig:hd209_var} and \ref{fig:hd189_var}) and this means that the planetary signal is contained in different components in each detector.  An optimal approach should adapt the number of components per detector based on their variance.
	
	Future work will consider the use of the algorithm presented here to analyse high-resolution observations taken by other instruments. These include CRIRES+ \citep{Follert2014},  GIANO-B, a high-dispersion spectrograph at TNG \citep{Oliva2012} which covers  $0.9$ to $2.5\ \mu$m with a resolution of (R=50,000); IRCS-SUBARU \citep{Kobayashi2000}, which uses a lower resolution (R=20,000) but covers a broader range (from $1$ to $5\ \mu$m) and CARMENES at Calar Alto Observatory \citep{Quirrenbach2014} with a spectral resolution up to 80,000 in the near-IR ($0.9-1.7\ \mu $m).
	
	\section*{Acknowledgments} \label{sec:akno}
	
	M.D and G.T. are supported by the European Research	Council (ERC) grant 617119 ExoLights. M.D. and G.M. are supported by the project "A Way to Other Worlds" funded by Italian Minister for University and Research. This work was also supported by STFC (ST/P000282/1).
	The authors wish to thank Dr. Antonino Petralia, Dr. Ingo Waldmann, Prof. Jonathan Tennyson, and Dr. Sergey Yurchenko for their input. \edit1{Finally, we thank the anonymous referee for the comments provided which helped to improve the quality of the manuscript.}
	
	\newpage
	
	{	\small
		\bibliographystyle{apj}
		\bibliography{HRS_bib.bib,HRS_bib2.bib}
	}	
	
\end{document}